\documentclass[article]{jss}

\usepackage{thumbpdf,lmodern}
\usepackage{comment}

\usepackage{framed}
\RequirePackage[OT1]{fontenc}
\RequirePackage{amsthm,amsmath,amsfonts,booktabs}
\RequirePackage[authoryear]{natbib}
\RequirePackage{bm,dsfont}
\usepackage{threeparttable}
\usepackage{comment}
\usepackage[T1]{fontenc}
\usepackage{changepage}
\usepackage[utf8]{inputenc}
\usepackage{pdflscape}
\usepackage{afterpage}
\usepackage{multirow}
\usepackage{tikz}
\usetikzlibrary{shapes.geometric, arrows, calc,
                positioning,
                quotes}
\tikzstyle{start} = [rectangle, rounded corners, 
minimum width=2cm, 
minimum height=1cm,
text centered, 
draw=black]

\tikzstyle{io} = [trapezium, 
trapezium stretches=true, 
trapezium left angle=70, 
trapezium right angle=110, 
minimum width=3cm, 
minimum height=1cm, text centered, 
draw=black, fill=blue!30]

\tikzstyle{process} = [rectangle, rounded corners,
minimum width=3in, 
minimum height=1cm, 
align = flush center, 
text width = 3.5in,
draw=black]

\tikzstyle{decision} = [circle, 
minimum width=1cm, 
minimum height=1cm, 
text centered, 
draw=black]

\tikzstyle{unobserved} = [circle, 
minimum width=1cm, 
minimum height=1cm, 
text centered,
dashed,
draw=black]

\tikzstyle{arrow} = [thick,->,>=stealth]

\numberwithin{equation}{section}
\theoremstyle{plain} 
\theoremstyle{plain} 
\theoremstyle{plain} 
\theoremstyle{plain} 
\theoremstyle{plain} \newtheorem{assumption}{{\sc Assumption}}

\theoremstyle{plain}

\theoremstyle{plain} 

\theoremstyle{plain}

\newcommand{\bX}{\bm{X}}
\newcommand{\bZ}{\bm{Z}}

\def\mis{\textup{mis}}
\def\obs{\textup{obs}}
\def\co{\textup{co}}
\def\nt{\textup{nt}}
\def\at{\textup{at}}
\def\df{\textup{df}}

\DeclareMathOperator\bE{\mathbb E} 

\newcommand{\indep}{\perp\!\!\!\perp}

\newcommand{\bW}{\mathbf{D}}
\newcommand{\bY}{\mathbf{Y}}

\newcommand{\be}{\begin{eqnarray}}
\newcommand{\ee}{\end{eqnarray}}
\newcommand{\bee}{\begin{eqnarray*}}
\newcommand{\eee}{\end{eqnarray*}}
\newcommand{\bi}{\begin{enumerate}[(i)]}
\newcommand{\ei}{\end{enumerate}}

\newcommand{\PS}{Principal Stratification }

\author{Bo Liu\\Duke University
   \And Fan Li  \\ Duke University}

\title{\proglang{PStrata}: An \proglang{R} Package for Principal Stratification}
\Plaintitle{PStrata: Principal Stratification in R}
\Shorttitle{PStrata: Principal Stratification in \proglang{R}}

\Abstract{
Post-treatment confounding is a common problem in causal inference, including special cases of noncompliance, truncation by death, surrogate endpoint, etc. Principal stratification \citep{FrangakisRubin02} is a general framework for defining and estimating causal effects in the presence of post-treatment confounding. A prominent special case is the instrumental variable approach to noncompliance in randomized experiments \citep{Angrist96}. Despite its versatility, principal stratification is not accessible to the vast majority of applied researchers because its inherent latent mixture structure requires complex inference tools and highly customized programming. We develop the \proglang{R} package \pkg{PStrata} to automatize statistical analysis of principal stratification for several common scenarios. \pkg{PStrata} supports both Bayesian and frequentist paradigms. For the Bayesian paradigm, the computing architecture combines R, C++, Stan, where R provides user-interface, Stan automatizes posterior sampling, and C++ bridges the two by automatically generating Stan code. For the Frequentist paradigm, \pkg{PStrata} implements a triply-robust weighting estimator. \pkg{PStrata} accommodates regular outcomes and time-to-event outcomes with both unstructured and clustered data. 
}

\Keywords{Causal inference, Compliers average causal effect, Instrumental variable, Principal stratification, Noncompliance, Truncation-by-death, Stan}
\Plainkeywords{JSS, style guide, comma-separated, not capitalized, R}

\Address{
  Bo Liu, Fan Li\\
  Department of Statistical Science\\ 
  Duke University\\ 
 Box 90251\\ 
  Durham, NC 27705, United States of America\\
  E-mail: \email{bo.liu1997@duke.edu}, \email{fl35@duke.edu}\\
  \\
  }

\begin{document}

\section[Introduction]{Introduction} \label{sec:intro}
Post-treatment confounding is a common problem in causal inference. Special cases include but not limited to noncompliance in randomized experiments, truncation by death, surrogate endpoint, treatment switching, and recruitment bias in cluster randomized trials. Principal stratification (PS) \citep{FrangakisRubin02} is a general framework for defining and estimating causal effects in the presence of post-treatment confounding. The instrumental variable approach to noncompliance in randomized experiments \citep{Angrist96} can be viewed as a special example of PS. Despite its versatility, PS is not accessible to most domain scientists because it requires complex modeling and inference tools such as mixture models and highly customized programming. We develop the \proglang{R} package \pkg{PStrata} to automatize statistical analysis of several most common scenarios of PS. \pkg{PStrata} supports both Bayesian and Frequentist inferential strategies. For Bayesian inference, \pkg{PStrata}'s computing architecture  combines R, C++, Stan, where R provides user-interface, Stan automatizes posterior sampling, and C++ bridges the two by automatically generating corresponding Stan code. For Frequentist inference, \pkg{PStrata} implements a triply-robust weighting estimator \cite{jiang2022}. \pkg{PStrata} accommodates regular outcomes and time-to-event outcomes with both unstructured and clustered data. 

This article reviews and illustrates the \pkg{PStrata} package. In Section \ref{sec:method}, we review the statistical framework of PS and its several most common special cases. Section \ref{sec:package} describes the main functions in \pkg{PStrata}. Section \ref{sec:illustrations} illustrates these functions with two simulated examples on noncompliance with both continuous and time-to-event outcomes. 

\section{Overview of Principal Stratification} \label{sec:method}

Before diving into the details of \pkg{PStrata}, we briefly introduce the basics of the PS framework. Post-treatment confounded variables lie on the causal pathway between the treatment and the outcome; they are potentially affected by the treatment and also affect the response. Since the landmark papers by \cite{Angrist96} and \cite{FrangakisRubin02}, a large literature on this topic has been developed. The post-treatment variable setting includes a wide range of specific examples, such as noncompliance \citep[e.g.][]{ImbensRubin97}, outcomes censored due to death \citep[e.g.][]{Rubin06, Zhang09}, surrogate endpoints \citep[e.g.][]{Gilbert08, jiang2016principal}. Below we first review the framework in the context of noncompliance in randomized experiments.

For unit $i\ (i=1,\ldots, N)$ in an \emph{iid} sample from a population, let $Z_i$ be the treatment assigned to ($1$ for treatment and $0$ for control), and $D_i$ be the treatment received ($1$ for treatment and $0$ for control). When $Z_i \neq D_i$, noncompliance occurs. Because $D$ occurs post assignment, it has two potential values, $D_i(0)$ and $D_i(1)$, with $D_i=D_i(Z_i)$. The outcome $Y_i$ also has two potential outcomes, $Y_i(0)$ and $Y_i(1)$. \citet{Angrist96} classified the units into four compliance types based on their joint potential status of the actual treatment, $S_i = (D_i(1), D_i(0))$: compliers $S_i=(1,0) = \co$, never-takers $S_i=(0,0) = \nt$,   always-takers $S_i=(1,1) =\at$, and defiers $S_i=(0,1) = \df$. This classification is  later generalized to principal stratification \citep{FrangakisRubin02} and the compliance types are special cases of principal strata. 

The key property of principal strata $S_i$ is that they are
not affected by the treatment assignment, and thus can be regarded
as a pre-treatment variable. Therefore, comparisons of $Y_i(1)$ and
$Y_i(0)$ within a principal stratum---the principal causal effects (PCEs)---have a causal interpretation:
$$
 \tau_u = \bE\{ Y_i(1) - Y_i(0) \mid S_i=s \} = \bE\{ Y_i(1)   \mid S_i=s \} - \bE\{  Y_i(0) \mid S_i=s\},
$$
for $u=\nt,\co,\at,\df$. The conventional causal estimand in the noncompliance setting is the intention-to-treat (ITT) effect, which is equivalent to the average causal effect of the assignment (ATE) and can be decomposed into the sum of the four PCEs:
$$
\textsc{ITT} = \bE\{ Y_i(1) - Y_i(0) \} = \sum_{s}   \pi_s \tau_s,
$$
where $\pi_s  = \Pr( S_i=s) $ is the proportion of the stratum $s$. 

\subsection{Randomized Experiments with Noncompliance}
Due to the fundamental problem of causal inference, individual compliance stratum $S_i$ is not observed. Therefore, additional assumptions are required to identify the stratum-specific effects. Besides randomization of $Z_i$, \cite{Angrist96} make two additional assumptions.

\begin{assumption}\label{as:monotonicity}
(Monotonicity): $D(1)\geq D(0)$, i.e. there is no $10$ or defiers stratum. 
\end{assumption}
Under monotonicity, there are only two strata--never-takers and compliers---in the \emph{one-sided noncompliance} setting (i.e. control units have no access to  treatment), and three strata---never-takers, compliers and always-takers---in the \emph{two-sided noncompliance} setting (i.e. control units also have access to treatment).

\begin{assumption}\label{as:ER}
(Exclusion restriction): for stratum $s$ in which $D(0)=D(1)$, i.e., $s=00, 11$, $E[Y_i(1)-Y_i(0)|S_i=s]=0$. 
\end{assumption}
In the context of noncompliance, exclusion restriction states that for never-takers and always-takers, the causal effects of the assignment $Z$ is 0. Exclusion restriction rules out the direct effect from the assigned treatment to the outcome, no mediated through the actual treatment. 

Under monotonicity and exclusion restriction, the complier average causal effect  is nonparametrically identifiable as
$$
\tau_{\co}
\equiv \bE\{ Y(1) - Y(0) \mid S = \co   \}
= \frac{  \bE(Y\mid Z=1)-\bE(Y\mid Z=0) }{  \bE(D\mid Z=1)-\bE(D\mid Z=0) } ,
$$ 
which is exactly the probability limit of the two-stage least squares estimator \cite{Angrist96}. Because under monotonicity, only compliers' actual treatments are affected by the assignment, $\tau_{\co}$ can be interpreted as the effect of the treatment.

\subsection{Truncation by Death}
``Truncation by death'' is a case where the outcome is not defined on some units whose intermediate variable is of certain value. The name comes from an original example, where treatment $Z$ is a medical invention, outcome $Y$ is quality of life at the end of the study, and $D$ is the survival status of the patient at the end of the study. Arguably quality of life is only defined for patients who are alive, and thus the outcome is ``truncated or censored by death" for patients who died. \cite{Rubin06} proposed to use \PS to address truncation by death. Here the four strata are sometimes named as ``never-survivor'', ``always-survivor'', ``compliers'' and ``defiers''. Treatment effect is well-defined  only in the ``always-survivor'' stratum, and is called ``survivor average causal effect (SACE)''. Monotonicity is usually still plausible, but exclusion restriction no longer applies. various another assumptions have been proposed in the literature.   


\subsection{Model-based analysis of PS}
We now describe the Bayesian inference of the IV setup, first outlined in \cite{ImbensRubin97}. Without any assumption, the observed cells of $Z$ and $D$ consist of a mixture of units from more than one stratum. For example, the units who are assigned to the treatment arm and took the treatment ($Z=1, D=1$) can be either always-takers or compliers. One has to disentangle the causal effects for different compliance types from observed data. Therefore, model-based inference here resembles that of a mixture model. In Bayesian analysis, it is natural to impute the missing label $S_i$ under some model assumptions. Specifically, to conduct Bayesian inference with posttreatment variables, besides the prior distribution for the parameter $\theta$, one need to specify two models:
\begin{enumerate}
    \item  S-model, i.e., principal strata model given covariates: $S_i\mid X_i$;
    \item  Y-model, i.e., outcome model given stratum, covariates and treatment: $Y_i\mid S_i, X_i, Z_i$ 
\end{enumerate}
A common model choice is a multinomial logistic regression for $S_i$ and a generalized linear model for $Y_i$. We can simulate the joint posterior distribution $\Pr(\theta, \bW^{\mis}\mid\bY^{\obs}, \bW^{\obs},\bZ, \bX)$ by iteratively imputing the missing $S$ from $\Pr(\bW^{\mis} \mid\bY^{\obs}, \bW^{\obs}, \bZ, \bX, \theta)$ and updating the posterior distribution of $\theta$ from $\Pr(\theta\mid\bY^{\obs}, \bW^{\obs}, \bW^{\mis}, \bZ, \bX)$. For more details, see \cite{ImbensRubin97, hirano00}.

\subsection{Frequentist approach under principal ignorability}
For the frequentist approach, \pkg{PStrata} implements the weighting-based multiply-robust estimator in \cite{jiang2022multiply}. The key assumption is principal ignorability \citep{JoStuart09}.
\begin{assumption}
(Principal Ignorability) $\E[Y(1)|s=11,X]=\E[Y(1)|s=10,X]$ and $E[Y(0)|s=00,X]=\E[Y(0)|s=10,X]$
\end{assumption}
Principal ignorability assumes away the mixture structure of principal stratification. Instead one can construct an inverse probability weighting based multiply-robust estimator. Three quantities are required. 
Define $\pi(X)=\Pr(Z=1|X)$ as the propensity score or the treatment probability, and $e_u(X)=\Pr(S=s|X)$ for $s=00, 01, 10, 11$ as the principal score. Let $\pi=\E[\pi(X)]=P(Z=1)$ and $e_s=\E[e_u(X)]=\Pr(S=s)$ be the marginalized propensity score and principal score. Also define $p_z(X)=\Pr(W=1|Z=z,X)$ for $z=0,1$, the probability of the intermediate variable conditional on the treatment and covariates. Under monotonicity, it is easy to see that
\[
e_{10}(X) = p_1(X)-p_0(X),\quad e_{11}(X)=p_0(X),\quad e_{00}(X)=1-p_1(X)
\]
and
\[
e_{10} = p_1-p_0,\quad e_{11}=p_0,\quad e_{00}=1-p_1
\]
where $p_z=\E[p_z(X)]$. The weighting-based estimator $\hat{\tau}^{mr}$ is a product of the inverse probability of the three scores: propensity score, principal score and the probability of the intermediate variable. If any two of the three models are correctly specified, then the estimator  $\hat{\tau}^{mr}$ is consistent. With principal ignorability, we still need monotonicity, but not necessarily exclusion restriction.  More details are given in \cite{jiang2022multiply}.


\section{Overview of the package} \label{sec:package}

The Bayesian inference consists of three main steps. First, define data generating models, which are equivalent to specifying the joint likelihood of all observed data given  parameters. Also, specify the prior distribution for these parameters. Second, draw samples from the posterior distribution of these parameters, which is obtained directly from the prior distribution and likelihood function. Third, calculate quantities of interest, which are functions of the parameters, with these posterior draws.

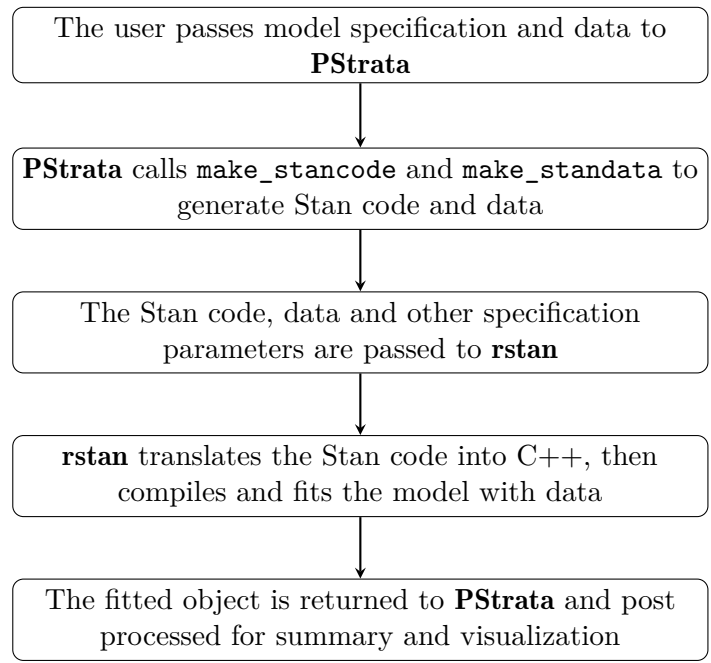
\begin{figure}[h]
\centering
\begin{tikzpicture}[node distance=0.75in]

\node (userinput) [process] {The user passes model specification and data to \pkg{PStrata}};
\node (makestan) [process, below of = userinput] {\pkg{PStrata} calls \code{make_stancode} and \code{make_standata} to generate Stan code and data};
\node (rstan) [process, below of = makestan] {The Stan code, data and other specification parameters are passed to \pkg{rstan}};
\node (sampling) [process, below of = rstan] {\pkg{rstan} translates the Stan code into C++, then compiles and fits the model with data};
\node (outcome) [process, below of = sampling] {The fitted object is returned to \pkg{PStrata} and post processed for summary and visualization};

\draw [arrow] (userinput) -- (makestan);
\draw [arrow] (makestan) -- (rstan);
\draw [arrow] (rstan) -- (sampling);
\draw [arrow] (sampling) -- (outcome);

\end{tikzpicture}
\caption{Overview of the model fitting process in \pkg{PStrata}}
\label{fig:overviewPStrata}
\end{figure}

The Bayesian component of the package \pkg{PStrata} is designed to follow these steps closely, which is illustrated in Figure \ref{fig:overviewPStrata}. First, the user specifies the data generating models and the prior distribution of the parameters in these models. As we shall see later, these data generating models can be  constructed by simply the S-model, Y-model and assumptions. Then, \pkg{PStrata} relies on the \pkg{Stan} language to draw samples from the posterior distribution. With user specification of the models and the dataset, \pkg{PStrata} automatically generates corresponding \pkg{Stan} code and \pkg{Stan} data. The code and data are processed by \pkg{Stan} to facilitate the posterior sampling process. Finally, these posterior samples are passed back to \pkg{PStrata} for post-process, such as calculation of estimands, summary and visualization, etc.

\begin{figure}[h]
\centering
\begin{tikzpicture}[node distance=1.5in]

\node (PSObject) [start] {PSObject};
\node (PSSample) [start, below of = userinput] {PSSample};
\node (PStrata) [start, right = 1in of $(PSObject.east)!0.5!(PSSample.east)$] {PStrata};
\node (PSOutcome) [start, right of = PStrata] {PSOutcome};
\node (PSContrast) [start, right of = PSOutcome] {PSContrast};

\draw [arrow] (PSObject) -- (PStrata);
\draw [arrow, dotted] (PSObject) -- (PSSample);
\draw [arrow] (PSSample) -- (PStrata);
\draw [arrow, dotted] (PStrata) -- (PSOutcome);
\draw [arrow, dotted] (PSOutcome) -- (PSContrast);

\end{tikzpicture}
\caption{Core modules designed in \pkg{PStrata}. Solid arrows indicate that the start-point module is called by the end-point module. Dotted arrows indicate that the outcome of the start-point module is passed to the end-point module as an argument.}
\label{fig:modulePStrata}
\end{figure}
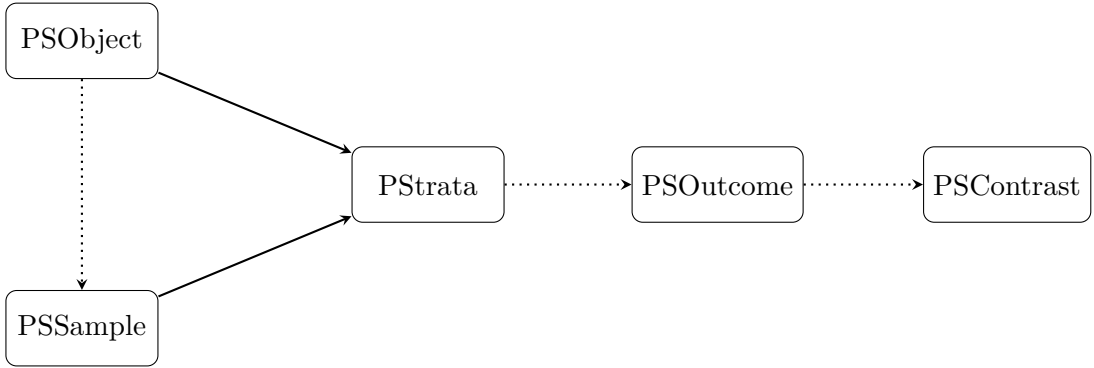

The main function in \pkg{PStrata} with the same name, \code{PStrata}, is designed to be an integrated module that bridges between the front-end and the back-end. On the front-end, users input model specification and data, and then receive posterior samples as the output. The sampling process in the back-end should be fully decoupled from the front-end so that users do not need to worry about sampling as they specify the models, which is unlikely possible for Gibbs samplers. \code{PStrata} incorporates two core submodules \code{PSObject} and \code{PSSample} to fulfill this requirement. The \code{PSObject} submodule interprets user input to generate a principal stratification object that contains essential information about model specification. The \code{PSSample} submodule calls \pkg{Stan} to generate posterior samples. These posterior samples are returned in \code{PStrata} and can be further passed to \code{PSOutcome} and \code{PSContrast} for calculating potential outcomes and causal effects. The relationship between these modules is illustrated in Figure \ref{fig:modulePStrata}.

In Section x.x, we introduce the details of model specification. In Section x.x, .....

\subsection{Model specification}

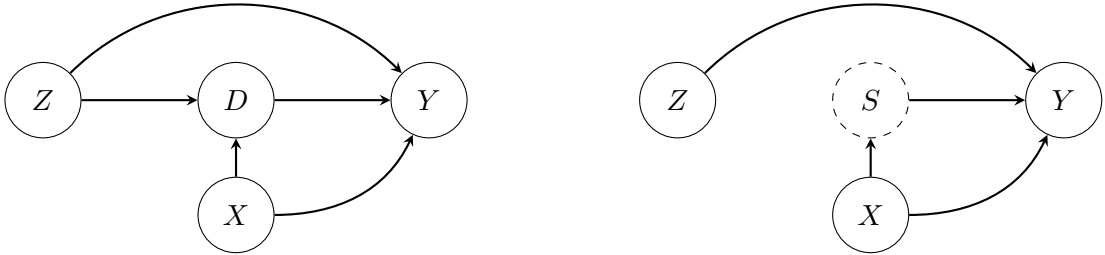
\begin{figure}[h]
\centering
\begin{tikzpicture}[node distance=1in]

\node (Z) [decision] {$Z$};
\node (D) [decision, right of = Z] {$D$};
\node (X) [decision, below = 0.2in of D] {$X$};
\node (Y) [decision, right of = D] {$Y$};

\draw [arrow] (Z) -- (D);
\draw [arrow] (X) -- (D);
\draw [arrow] (D) -- (Y);
\draw [arrow] (Z) to [out=45,in=135] (Y);
\draw [arrow] (X) to [out=0,in=245] (Y);

\end{tikzpicture}
\hspace{2cm}
\begin{tikzpicture}[node distance=1in]

\node (Z) [decision] {$Z$};
\node (S) [unobserved, right of = Z] {$S$};
\node (X) [decision, below = 0.2in of D] {$X$};
\node (Y) [decision, right of = D] {$Y$};

\draw [arrow] (S) -- (Y);
\draw [arrow] (X) -- (S);
\draw [arrow] (Z) to [out=45,in=135] (Y);
\draw [arrow] (X) to [out=0,in=245] (Y);

\end{tikzpicture}
\caption{Graph illustration of principal stratification. The left panel shows that post-treatment covariate $D$ lies between the causal pathway from treatment $Z$ to outcome $Y$. The right panel shows that the unobservable principal stratum $S$ separates $D$ from the causal pathway from $Z$ to $Y$.}
\label{fig:PSCausalGraph}
\end{figure}

Typical datasets used for principal stratification consist of the treatment $Z$, the post-treatment covariate $D$, the outcome $Y$ and covariates $X$. To better convey to main idea, we assume that both $Z$ and $D$ are binary variables, but the idea can be readily extended to the case of non-binary but discrete $Z$ and $D$, and the case of multiple post-treatment covariates (when $D$ is a vector). Since $D$ is observed after treatment, controlling $D$ leads to biased estimation of the causal effect, as it breaks randomization on the treatment $Z$. However, the principal stratum $S$, defined by the potential values of $D$ under each treatment, can be viewed as a pre-treatment variable \citep{ImbensRubin97}. When the principal stratum $S$ is fixed, no variables lie on the causal pathway from $Z$ to $Y$; thus the stratum-specific causal effect can be correctly estimated (Figure \ref{fig:PSCausalGraph}).

If the principal stratum $S$ were known, the Y-model $\Pr(Y \mid X, Z, S)$ would be sufficient to estimate any causal estimand. However, as $S$ is defined as the potential values of $D$, only one realization of them can be observed.  Hence, a model for principal stratum $\Pr(S \mid X)$ is necessary to estimate $S$. Let $\mathcal{S}$ denote the set of all possible principal strata. The data generation process can be expressed as follows.
\begin{align}
    \Pr(Z, D, Y \mid X) &= \sum_{s\in\mathcal{S}} \Pr(Z\mid X)\Pr(S = s\mid Z, X) \Pr(D \mid S = s, Z, X)\Pr(Y \mid D, S = s, Z, X) \label{eq:fulldecomposition} \\
    &\propto\sum_{s\in\mathcal{S}: D = D(s, Z)} \Pr(S = s\mid X)\Pr(Y\mid S = s, Z, X). \label{eq:mixturemodel}
\end{align}
The first term $\Pr(Z \mid X)$ in \eqref{eq:fulldecomposition} is a constant for randomized trials. The third term $\Pr(D \mid S = s, Z, X)$ is deterministic, and controls the set of principal strata over which the summation is taken. $D$ in the last term $\Pr(Y \mid D, S = s, Z, X)$ drops out as it has no additional information given $Z$ and $S$. 

It is shown in \eqref{eq:mixturemodel} that the data generating model can be expressed as a mixture model involving only the S-model and Y-model, and our aim is to disentangle such mixture from the observed data. Depending on the real application, we may impose additional assumptions to help identify the components in the mixture model. The first assumption is monotonicity which rules out defiers. In some cases when the treatment cannot be accessed by the control group, we may also rule out always-takers. In general, this assumption defines the set $\mathcal{S}$ of all strata that we include in \eqref{eq:fulldecomposition}. The second assumption is the exclusion-restriction (ER). By assuming ER for stratum $s$, we exclude the direct causal pathway from $Z$ to $Y$ given $S$, and consequently, 
\begin{equation}
    \Pr(Y \mid S = s, Z = z_1, X) = \Pr(Y \mid S = s, Z = z_2, X),\quad\text{if $D(s, z_1) = D(s, z_2)$}.
\end{equation}

\pkg{PStrata} offers the \code{PSObject} function to specify the S-model and Y-model, put assumptions on monotonicity and ER, and impose prior distributions on parameters that appear in the models. The function has the following declaration:
\begin{Code}
PSObject(S.formula, Y.formula, Y.family, data, strata, ER, 
+    prior_intercept, prior_coefficient, prior_sigma,
+    prior_alpha, prior_lambda, prior_theta, survival.time.points)
\end{Code}

We will discuss in detail these arguments in subsequent sections, with the exception of \code{data} and \code{survival.time.points}. The \code{data} argument simply provides the dataset to fit the model. The \code{survival.time.points} argument is only used for survival models, which will be discussed later in Section x.x.

\subsubsection{S-model and Y-model}

In \pkg{PStrata}, we specify the S-model $p(S_i\mid X_i)$ as a multinomial model, 
\begin{equation}
    \Pr(S_i = s \mid X_i; \{\beta_s\}) \propto \exp(X_i' \beta_s). \label{eq:S-model}
\end{equation}
When an intercept is included in the model, the $X_i$ in \eqref{eq:S-model} can be understood as an augmented vector with the constant 1.
 To assure identifiability, a reference stratum $s_0 \in \mathcal{S}$ is chosen and $\beta_{s_0}$ is fixed to zero. The S-model is equivalently expressed by 
\begin{equation}
    \Pr(S_i = s \mid X_i; \{\beta_s\}) = \frac{\exp(X_i' \beta_{s})}{\sum_{s'\in \mathcal{S}} \exp(X_i' \beta_{s'})}. \label{eq:S-model-equiv}
\end{equation}

We specify the Y-model as a generalized linear model, which is flexible enough in modeling various types of data. The Y-model can be mathematically written as
\begin{equation}
    Y_i\mid S_i = s, X_i, Z_i = z \sim F(\mu_{sz}(X_i), \theta_{sz}), \label{eq:Y-model-family} \\
\end{equation}
where $F$ is a distribution often referred to as the family of the generalized linear model. The mean of this distribution is related to the linear combination of $X_i$ via a link function $g$: 
\begin{equation}
    \mu_{sz}(X_i) := \E[Y_i\mid S_i = s, X_i, Z_i = z, \gamma_{sz}] = g^{-1}(X_i'\gamma_{sz}), \label{eq:Y-model-link}
\end{equation}
where $\{\gamma_{sz}\}$ are parameters to be estimated. Some family $F$ may have additional parameters apart from the mean, such as the standard deviation parameter in Gaussian models and the shape parameter in Gamma and inverse Gaussian models. When these parameters exists, we assume they are independent of the covariates. However, we allow these parameters to depend on the stratum $s$ and treatment $z$, as denoted by $\theta_{sz}$ in \eqref{eq:Y-model-family}.

The S-model and Y-model are specified with arguments \code{S.formula}, \code{Y.formula} and \code{Y.family}. An (incomplete) example is given as follows.
\begin{Code}
    ...
    S.formula = Z + D ~ X,
    Y.formula = Y ~ X,
    Y.family = binomial(link = "logit"),
    ...
\end{Code}

The terms before the \code{~} symbol defines the variables essential to the pricipal stratification model. Specifically, \code{S.formula} defines the treatment variable $Z$ and the post-treatment covariate $D$ concatenated by the \code{+} sign, and \code{Y.formula} defines the response variable $Y$. When multiple post-treatment covariates exists, they can be concatenated in order like \code{S.formula = Z + D1 + D2 ~ X}. However, we do not allow multiple treatment variables to avoid ambiguity. In cases where multiple treatment variables exists, they might be recoded into a single variable with multiple treatment arms.

The terms after the \code{~} symbol defines the covariates that are included in \eqref{eq:S-model} and \eqref{eq:Y-model-link}, with the syntax exactly matching that of \code{lm()}. To recall the basic syntax, predictors are separated by \code{+} as in \code{Y ~ X1 + X2}. By default, the intercept term is included; either \code{Y ~ X1 + X2 + 0} or \code{Y ~ X1 + X2 - 1} removes the intercept. Each predictor can be a raw covariate (\emph{e.g.} \code{X1}), a transformed covariate (\emph{e.g.} \code{I(X1\^{}2)}) or the interaction of multiple covariates (\emph{e.g.} \code{X1:X2}).

The \code{Y.family} argument specifies the family and link of the generalized linear model following the convention of \code{glm()}. The families and link functions supported in \pkg{PStrata} are listed in Table \ref{tab:supportedfamilies}.

\begin{table}[h]
    \centering
    \begin{tabular}{ccc}
    \toprule
    Family & Link Functions & $f(y; \mu, \theta)$\\ 
    \midrule
    \code{gaussian} & \code{identity}, \code{log}, \code{inverse} & $\frac{1}{\sqrt{2\pi\sigma^2}}\exp\left(-\frac{-(y-\mu)^2}{2\sigma^2}\right)$ \\
    \code{binomial} & \code{logit}, \code{probit}, \code{cauchit}, \code{log}, \code{cloglog} & $\mu^y(1-\mu)^{1-y}$\\
    \code{Gamma} & \code{identity}, \code{log}, \code{inverse} & $\frac{(\alpha/\mu)^\alpha}{\Gamma(\alpha)}y^{\alpha - 1}\exp(-\frac{\alpha y}{\mu})$\\
    \code{poisson} & \code{identity}, \code{log}, \code{sqrt} & $\mu^y e^{-\mu}/(y!)$ \\
    \code{inverse.gaussian} & \code{1/mu\^{}2}, \code{inverse}, \code{identity}, \code{log}&  $\sqrt{\frac{\lambda }{2\pi y^3}}\exp\left(-\frac{\lambda(y-\mu)^2}{2y\mu^2}\right)$\\
    \bottomrule
    \end{tabular}
    \caption{Supported families, link functions and corresponding density/mass functions expressed by the mean $\mu$ and additional parameter $\theta$.}
    \label{tab:supportedfamilies}
\end{table}

\subsubsection{Assumptions}
In the context of binary non-compliance, there are two important assumptions: monotonicity (Assumption \ref{as:monotonicity}) and exclusion restriction (Assumption \ref{as:ER}). Monitonicity rules out the existence of defiers and thus all strata to be considered are always-takers, never-takers, and compliers. This can be specified by \code{strata = c(at = "11", nt = "00", co = "01")}.  The exclusion restriction (ER) assumes away direct effects of $Z$ on $Y$ not through $D$ for certain strata, which is specified by a binary logical vector with the same length as \code{strata}, indicating whether ER is assumed for each stratum. For example, \code{ER = c(True, True, False)} assumes ER for both the always-takers and the never-takers, but not for the compliers. For simplicity, we introduce a short-hand specification for both assumptions by adding an asterisk (\code{*}) after the group where ER is assumed. As an example, the assumptions mentioned above can be specified by \code{strata = c(at = "11*", nt = "00*", co = "01")}, and the \code{ER} argument can be left blank.

When multiple intermediate variables exist, the principal strata are defined by the potential values of each intermediate variable. To specify such a stratum, use \code{|} to separate the potential values of each intermediate variable. For example, the principal stratum defined by two intermediate variables $D_1(0) = D_1(1) = 0$ and $D_2(0) = D_2(1) = 1$ can be represented by \code{"00|11"}.

\subsubsection{Priors}

The S-model and Y-model are both parametric models. These parameters include intercepts ($\alpha_s$ and $\gamma_{sz}$), coefficients ($\beta_s$ and $\eta_{sz}$), and other parameters specific to the family of the Y-model. As Table \ref{tab:supportedfamilies} shows, the \code{gaussian} family introduces a parameter $\sigma$ that represents the standard deviation of the error, the \code{Gamma} family introduces the shape parameter $\alpha$, and the \code{inverse.gaussian} family introduces the shape parameter $\lambda$. We refer to these parameters as \code{sigma}, \code{alpha} and \code{lambda} acoording to convention.

\pkg{PStrata} provides flexibility on the specification of prior distribution on these parameters, including the intercept, coefficients and model-specific parameters. The prior distribution of intercepts and coefficients can be specified in \code{prior_intercept} and \code{prior_coefficient} respectively. When additional parameters exists in Y-model, the prior distributions of such variables are defined by \code{prior_sigma}, \code{prior_alpha} or \code{prior_lambda}, according to the name of the additional parameter. The full list of supported prior distributions is provided in Table \ref{tbl::prior}.

By default, \code{prior_intercept} is a flat (improper) distribution over the whole real line, \code{prior_coefficient} is a standard gaussian distribution $\mathcal{N}(0, 1)$, and \code{prior_sigma}, \code{prior_alpha} and \code{prior_lambda} are standard inverse-gamma distribution $\mathcal{IG}(1, 1)$.

\begin{table}[h]
\centering
\begin{tabular}{ccc}
    \toprule
    Domain & Prior distribution & Specification \\
    \midrule
    \multirow{6}{*}{$(-\infty, \infty)$} & Uniform (improper) & \code{prior_flat()} \\
    & Normal & \code{prior_normal(mean = 0, sigma = 1)} \\
    & Student $t$ & \code{prior_t(mean = 0, sigma = 1, df = 1)} \\
    & Cauchy & \code{prior_cauchy(mean = 0, sigma = 1)} \\
    & Double exponential & \code{prior_lasso(mean = 0, sigma = 1)} \\
    & Logistic & \code{prior_logistic(mean = 0, sigma = 1)} \\
    \midrule
    \multirow{6}{*}{$(0, \infty)$} & Chi squared & \code{prior_chisq(df = 1)} \\
    & Inverse Chi squared & \code{prior_inv_chisq(df = 1)} \\
    & Exponential & \code{prior_exponential(beta = 1)} \\
    & Gamma & \code{prior_gamma(alpha = 1, beta = 1)} \\
    & Inverse Gamma & \code{prior_inv_gamma(alpha = 1, beta = 1)} \\
    & Weibull & \code{prior_weibull(alpha = 1, sigma = 1)} \\
    \bottomrule
    \end{tabular}
    \caption{List of prior distributions.}
    \label{tbl::prior}
\end{table}

\subsection{Other data types}

\subsubsection{Survival data}

Survival data refer to the type of data where the outcome is subject to right censoring. Suppose each subject has a true event time $T_i$ and censoring time $C_i$, but they cannot be observed due to censoring. The observed data are the censored event time $Y_i = \min\{T_i, C_i\}$ and the event indicator $\delta_i = \mathbf{1}(T_i < C_i)$. 
According to the data generating process, we assume that there is an underlying distribution $p(T_i, C_i \mid X_i, S_i, Z_i)$, which in turn decides the distribution of observed quantities $p(Y_i, \delta_i \mid X_i, S_i, Z_i)$. When the censoring is conditionally ignorable, \emph{i.e.}, $T_i \indep C_i \mid \{X_i, S_i, Z_i\}$, it turns out that the distribution of $T_i$ and $C_i$ can be separated from the likelihood $p(Y_i, \delta_i \mid X_i, S_i, Z_i)$:
\begin{align}
    p(Y_i, \delta_i = 1 \mid X_i, S_i, Z_i) &= p(T_i \geq Y_i\mid S_i, Z_i, X_i)p(C_i = Y_i \mid  S_i, Z_i, X_i), \\
    p(Y_i, \delta_i = 0 \mid X_i, S_i, Z_i) &= p(T_i = Y_i\mid S_i, Z_i, X_i)p(C_i \leq Y_i \mid  S_i, Z_i, X_i).
\end{align}
As we usually only care out the true event time $T_i$, the terms regarding the censoring time $C_i$ drops out from the likelihood, and only specification of the distribution for $T_i$ is needed.

Two widely used models for $T_i$ are the Cox proportional hazard model \citep{Cox72} and the accelerated failure time (AFT) model \citep{Wei92AFT}. The Cox proportional hazard model assumes that the hazard function should be
\begin{equation}\label{eq:Cox}
    h(t; S_i = s, Z_i = z, X_i) = h_0(t; s, z)\exp(X_i'\beta_{s,z}),
\end{equation}
where $h_0(t; s, z)$ is the baseline hazard function. In Bayesian inference, a parametric form of $h_0$ is desirable. A common form is the Weibull model \citep{Abrams96}: $h_0(t; s, z) = \exp(\alpha_{s, z})t^{\varphi_{s, z} - 1}$, $\alpha_{s,z}\in\mathbb{R}$, $\varphi_{s,z} > 0$,  under which the the Cox model \ref{eq:Cox} takes the parameteric form (under reparameterization)
\begin{equation}
    h(t; S_i = s, Z_i = z, X_i) = t^{\exp(\theta_{s,z}) - 1}\exp(X_i' \beta_{s,z}). \label{eq:Weibull-Cox}
\end{equation}
It can be then derived from the hazard function that 
\begin{align}
    p(T_i \geq Y_i\mid S_i = s, Z_i = z, X_i) &= \exp\left\{-t^{\exp(\theta_{s,z})}\exp(-\theta_{s,z} + X_i'\beta_{s,z})\right\}, \label{eq:surv-prob-Cox}\\
    p(T_i = Y_i\mid S_i = s, Z_i = z, X_i) &= h(t; s, z, X_i) p(T_i \geq Y_i\mid S_i = s, Z_i = z, X_i).
\end{align}
The AFT model assumes that $\log(T_i) = X_i'\beta + \epsilon_i$, where $\epsilon_i$ follows a given distribution. Commonly, we assume that $\epsilon_i \sim \mathrm{N}(0, \sigma^2)$. Then, it can be computed that
\begin{align}
    p(T_i \geq Y_i\mid S_i = s, Z_i = z, X_i) &=  1 - \Phi\left(\frac{\log Y_i - X_i'\beta_{s,z}}{\sigma_{s,z}}\right), \label{eq:surv-prob-AFT}\\
    p(T_i = Y_i\mid S_i = s, Z_i = z, X_i) &= \phi\left(\frac{\log Y_i - X_i'\beta_{s,z}}{\sigma_{s,z}}\right),
\end{align}
where $\Phi$ and $\phi$ are respectively the cumulative density function and probability density function of standard normal distribution.

\pkg{PStrata} allows fitting principal stratification models on survival data under the conditional ignorability assumption. Running models on survival data is similar to the normal data except some modification. First, one needs to include both censored event time $Y$ and the event indicator $\delta$ when specifying the Y-model. For example, \code{Y.formula = Y + delta ~ X} specifies a Y-model where both \code{Y} and \code{delta} appear on the left hand side of the formula. Second, the \code{Y.family} should be \code{survival(method = "Cox")} for the Weibull-Cox model or \code{survival(method = "AFT")} for the AFT model. Third, since the fitted outcome - the survival probability at time $t$ - is a continuous curve of $t$, one needs to specify a set of time points where the survival probability is calculated. One either specify as a vector the customized time points to the parameter \code{survival.time.points}, or simply provide a positive integer indicating the number of time points, and the time points will be uniformly chosen from 0 to the 90\% quantile of the observed event time.

\subsubsection{Multi-level data}

In previous sections, we assume that the units are independent samples from the population. In some applications, this assumption might be implausible when natural clusters exist, such as schools, hospitals and regions. For example, in a trial conducted at multiple medical centers, patients from the same medical center may have correlated outcomes. Hierarchical models are commonly used to model such data, by assuming that each cluster has a cluster-specific effect (random effect) drawn from some unknown distribution, and within each cluster, the outcome of the units follow some distribution related to the cluster-specific effect. The hierarchical structure is well-fit for Bayesian analyses, and \pkg{PStrata} allows specification of multi-level structure in both S-model and Y-model.

We introduce the most common setting where only one level of clusters exists, although \pkg{PStrata} provides flexibility to model various cluster strucures. Let $j = 1, \dots, J$ denote the clusters, and let $C_{i}$ denote the cluster which subject $i$ belongs to. Additional to the individual covariates $X_i$, let $V_j$ be cluster covariates of cluster $j$. Assuming the random effects are additive, the multi-level S-model assumes that
\begin{equation}
    \Pr(S_i = s\mid X_i, V_j, C_i = j) \propto \exp(X_i'\beta_s + V_j'\widetilde{\beta}_s + \eta_j), \label{eq:S-model-re}
\end{equation}
where $\eta_j \sim \mathrm{N}(0, \sigma^2)$ are cluster random effects. Similarly, the mean of the Y-model is specified as 
\begin{equation}
    \E(Y_i\mid Z_i = z, X_i, V_j, C_i = j)= g^{-1}(X_i' \gamma_{sz} + V_j' \widetilde{\gamma}_{sz} + \xi_{sz, j}), \label{eq:Y-model-re}
\end{equation}
where $\xi_{sz, j} \sim \mathrm{N}(0, \tau_{sz}^2)$ are the cluster random effects.

\pkg{PStrata} supports including these random effects when clusters exist, by specifying these effects in \code{S.formula} and/or \code{Y.formula} with the same syntax as in \code{lme4::lmer()}. The syntax to include the above-mentioned additive random effects is to include a term \code{(1 | C)}, where \code{C} is the cluster id. For example, the Y-model mentioned above can be specified as \code{Y.formula = Y ~ X + V + (1 | C)}, where \code{C} is a vector indicating the cluster id each unit belongs to.

The additive assumption for random effects can be relaxed to allow the coefficients $\beta_s$, $\widetilde{\beta}_s$, $\gamma_{sz}$, $\tilde{\gamma}_{sz}$ to vary across cluster. For example, a random effect on the coefficient $\beta_s$ will modify the S-model as 
\begin{equation}
    \Pr(S_i = s\mid X_i, V_j, C_i = j) \propto \exp(X_i'(\beta_s + \omega_j) + V_j'\widetilde{\beta}_s + \eta_j),
\end{equation}
where both $\eta_j$ and $\omega_j$ are random effects drawn from distributions $\eta_j \sim \mathrm{N}(0, \sigma^2), \omega_j \sim \mathrm{N}(0, \kappa_j^2)$. In the literature, $\eta_j$ is commonly referred to as random intercept, and $\omega_j$ is commonly referred to as random slope.

In \pkg{PStrata}, the above model can be specified as \code{S.formula = S ~ X + V + (X | C)}, where the term \code{(X | C)} specifies a random slope on $X$ and an (implicit) random intercept according to the cluster $C$. To remove the implicit random intercept, one should specify the model as \code{S.formula = S ~ X + V + (X - 1 | C)} or \code{S.formula = S ~ X + V + (X + 0 | C)}. More details on the syntax can be found in the documentation of  \code{lme4::lmer()}.

\subsection{Outcome and contrast}

\pkg{PStrata} calls \pkg{stan} to draw samples from the posterior distribution of the parameters. These posterior samples can be used to calculate summary statistics, such as the stratum probability, stratum-specific potential outcomes and their contrasts.

For each parameter, we denote the sample drawn in each iteration by a superscript in parenthesis. For example, the samples drawn from the posterior distrbution $\alpha$ are denoted by $\alpha^{(1)}, \alpha^{(2)}, \dots, \alpha^{(B)}$, where $B$ is the number of posterior draws.

\subsubsection{Stratum probability}
Recall the multinomial logistic S-model \eqref{eq:S-model}. In iteration $k$, the estimated probability of unit $i$ being in stratum $s$ can be obtained by plugging in the posterior draws from that iteration,
\begin{equation}
    p_{is}^{(k)} := p(S_i = s \mid X_i) = \frac{\exp(\beta_s^{(k)}{}' X_i)}{\sum_{\varsigma \in \mathcal{S}}\exp(\beta_{\varsigma }^{(k)}{}' X_i)}.
    \label{eq:stratum-probability-estimate}
\end{equation}
By calculating this quantity in each iteration, we obtain a sequence of estimated stratum probability $p_{is}^{(1)}, \dots, p_{is}^{(B)}$, which can be viewed as drawn from the posterior distribution of $p_{is}$. With these posterior samples, \pkg{PStrata} calculates and reports the expectation, the standard deviation and the quantiles of the estimated stratum probabililty.

\subsubsection{Outcome}
Let $\mu_{sz}$ denote the average potential outcome for stratum $s$ under treatment $z$ which is defined as 
\begin{align}
    \mu_{sz} &= \E[Y(z)\mid S = s] \nonumber\\
    &= \E[\E[Y(z)\mid S = s, X] \mid S = s] \nonumber \\
    &= \int \E[Y(z)\mid S = s, X = x] \frac{\Pr(S = s\mid X = x)}{\Pr(S = s)}\mathbb{P}(\mathrm{d}x) \nonumber \\
    &= \frac{\int \E[Y(z)\mid S = s, X = x] \Pr(S = s\mid X = x)\mathbb{P}(\mathrm{d}x)}{\int \Pr(S = s\mid X = x)\mathbb{P}(\mathrm{d}x)}.
\end{align}
To estimate such a quantity, we approximate the measure $\mathbb{P}(\mathrm{d}x)$ by the empirical distribution of $X$ in the data, and thus both integrals reduce to summation over each unit. Within the integrals, the first term $\E[Y(z)\mid S = s, X= x]$ is given by the Y-model \eqref{eq:Y-model-link}, 
\begin{equation}
    \E[Y_i(z)\mid S_i = s, X_i] = g^{-1}(\gamma_{sz}'X_i).
\end{equation}
The term $\Pr(S = s\mid X = x)$ is exactly the stratum probability $p_{is}$ defined in the previous section. This suggests estimating the potential outcome $\mu_{sz}$ by 
\begin{align}
    \mu_{sz} \approx \frac{\sum_{i=1}^n g^{-1}(\gamma_{sz}'X_i) p_{is}}{\sum_{i=1}^n p_{is}}.
\end{align}
By plugging in the posterior samples of the parameter $\gamma_{sz}$ and the stratum probability $p_{is}$, we obtain posterior draws $\mu_{sz}^{(1)}, \dots, \mu_{sz}^{(B)}$, where 
\begin{align}
    \mu_{sz}^{(k)} = \frac{\sum_{i=1}^n g^{-1}(\gamma_{sz}^{(k)}{}'X_i) p_{is}^{(k)}}{\sum_{i=1}^n p_{is}^{(k)}}.
\end{align}
The expectation, standard deviation and quantiles of the estimated potential outcomes can be obtained directly from these poterior draws.

\subsubsection{Contrast}

Many statistically meaningful estimands can be represented as contrasts of the potential outcomes. For example, the SATE for stratum $s$ is the difference between two potential outcomes: 
\begin{equation}
    \tau_s = \E[Y(1) - Y(0)\mid S = s] = \mu_{s1} - \mu_{s0}.
\end{equation}
The SATE can be further compared between strata $s$ and $s'$ by 
\begin{equation}
    \Delta_{s, s'} = \tau_s - \tau_s' = (\mu_{s1} - \mu_{s0}) - (\mu_{s'1} - \mu_{s'0}) = (\mu_{s1} - \mu_{s'1}) - (\mu_{s0} - \mu_{s'0}).
\end{equation}

To obtain the posterior distribution of these contrasts, the posterior samples of potential outcomes $\mu_{sz}^{(k)}$ are in place of $\mu_{sz}$, resulting in a sequence of contrasts. These values can be viewed as samples from the posterior distribution of the contrast, and expectation, standard deviation and quantiles can be readily obtained.

\pkg{PStrata} allows contrasts between either strata or treatment arms, and they can be nested in arbitrary order, which guarantees flexibility in defining various meaningful contrasts. As an example, $\tau_s$ is the contrast between treatment arms, and $\Delta_{s, s'}$ is the contrast between treatment arms followed by a contrast between strata, or in the reverse order.

\subsubsection{Survival data}

For survival data, the most common causal estimand is the survival probability, defined by $\Pr(T_i > t)$. Let $\mu_{sz}(t)$ denote the average survival probatility at time $t$ for stratum $s$ under treatment $z$:
\begin{align}
    \mu_{sz}(t) &= \Pr(T > t \mid S = s, Z = z) \\
    &= \frac{\int \Pr(T > t \mid S = s, Z = z, X = x)\Pr(S = s\mid X = x) \mathbb{P}(\mathrm{d}x)}{\int \Pr(S = s\mid X = x) \mathbb{P}(\mathrm{d}x)}.
\end{align}
For any given $t$, the term $\Pr(T > t\mid S = s, Z = z, X = x)$ can be calculated by \eqref{eq:surv-prob-Cox} under Cox model or \eqref{eq:surv-prob-AFT} under AFT model. Approximating the integral using summation over the empirical distribution of $X$, 
\begin{equation}
    \mu_{sz}(t)
    = \frac{\sum_{i=1}^n \Pr(T_i > t \mid S_i = s, Z_i = z, X_i)p_{is}}{\sum_{i=1}^n p_{is}}. \label{eq:outcome-survival-data}
\end{equation}
In \pkg{PStrata}, direct calculation of $\mu_{sz}$ as a function is not plausible. Instead, a grid of time points $t_1, \dots, t_m$ is pre-specified, and $\mu_{sz}$ is evaluated over these time points, $\mu_{szj} := \mu_{sz}(t_j)$. The posterior samples of parameters and $p_{is}$ are plugged into \eqref{eq:outcome-survival-data} to obtain the expectation, standard deviation and quantiles of $\mu_{szj}$. In addition to strata and treatment arms, contrasts between different time points are also supported for survival data.

\subsubsection{Multi-level data}

We introduce the calculation of outcome and contrast for multi-level data under S-model \eqref{eq:S-model-re} and Y-model \eqref{eq:Y-model-re}, but the idea can be easily extended to more complex model forms. Within each iteration, we not only sample coefficients from the posterior distribution, but also sample cluster random effects. Specifically, let $\beta_s^{(k)}$, $\widetilde{\beta}_s^{(k)}$ be the coefficients and let $\eta_j^{(k)}$ be the random effect sampled in the $k$-th iteration. Moreover, let $C_i$ denote the cluster that subject $i$ belongs to. Then, the stratum probability can be estimated by 
\begin{equation}
    p_{is}^{(k)} = \frac{\exp(X_i' \beta_s^{(k)} + V_{C_i}'\widetilde{\beta}_s^{(k)} + \eta_{C_i}^{(k)})}{\sum_{\varsigma \in \mathcal{S}}\exp(X_i' \beta_\varsigma^{(k)} + V_{C_i}'\widetilde{\beta}_\varsigma^{(k)} + \eta_{C_i}^{(k)})}.
\end{equation}
Similarly, the average potential outcome can be estimated by 
\begin{equation}
    \mu_{sz}^{(k)} = \frac{\sum_{i=1}^n g^{-1}(X_i'\gamma_{sz}^{(k)} + V_{C_i}'\widetilde{\gamma}_{sz}^{(k)} + \xi_{sz, C_i}^{(k)}) p_{is}^{(k)}}{\sum_{i=1}^n p_{is}^{(k)}},
\end{equation}
where $\gamma_{sz}^{(k)}$, $\widetilde{\gamma}_{sz}^{(k)}$ and $\xi_{sz, C_i}^{(k)}$ are samples from the posterior distribution obtained in the $k$-th iteration. The posterior distribution of the outcome and contrasts, summarized by the expectation, standard deviation and quantiles can be immediately obtained from these samples.

\section{Case study: Effect of flu vaccination on hospital visits} \label{sec:illustrations}

We illustrate the use of \pkg{PStrata} under the normal non-compliance setting in a case study that estimates the effect of influenza vaccine with data first studied by \cite{McDonald92flu} and reanalyzed by \cite{hirano00}. In the study, physicians were randomly chosen to receive a letter that encouraged them to vaccinate their patients for flu. The treatment of interest is the actual vaccination status of the patients, and the outcome is a binary variable indicating whether they have hospital visits for flu-related reasons. For each patient, some covariates are observed, among which we include age and chronic obstructive pulmonary disease (COPD) rates in the S-model and Y-model as in \cite{hirano00}.

In this case, the randomized treatment $Z_i$ is the encouragement indicator, and the outcome $Y_i$ is the indicator for hospital visits. The actual vaccination status, $D_i$, is a post-randomization intermediate variable. The conventional ITT approach estimates the causal effect of $Z_i$ on $Y_i$, which estimates the effectiveness of the encouragement, not the efficacy of the vaccination itself. Using the principal stratification framework, we define the principal strata by the potential vaccination status $S_i = (D_i(0), D_i(1))$, \emph{i.e.}, the always-vaccinated $S_i = (1, 1)$, the never-vaccinated $S_i = (0, 0)$, compliant-vaccinated $S_i = (0, 1)$ and defiant-vaccinated $S_i = (1, 0)$. Since flu vaccination can be taken regardless of the encouragement and is not mandate, there can be patients that are always-vaccinated or never-vaccinated. We assume that there are no defiant-vaccinated patients, but there are patients in all other three strata. Furthermore, we assume that for always-vaccinated and never-vaccinated patients, encouragement on vaccination itself does not have any direct effect on flu-related hospital visits. We are interested in estimating the causal effect of $Z_i$ among the compliant-vaccinated patients, as such causal effect estimates the efficacy of vaccination for this subpopulation.

\subsubsection{Create Principal Stratification Object (PSObject)}

\begin{CodeChunk}
\begin{CodeInput}
R> PSobj <- PSObject(
+    S.formula = encouragement + vaccination ~ age + copd,
+    Y.formula = hospital ~ age + copd,
+    Y.family = binomial(link = "logit"),
+    data = flu,
+    strata = c(n = "00*", c = "01", a = "11*"),
+    prior_intercept = prior_normal(0, 1),
+    prior_coefficient = prior_normal(0, 1)
+  )
\end{CodeInput}
\end{CodeChunk}

In the above code snippet, we created a principal stratification object \code{PSobj} by specifying the S-model and Y-model, providing the dataset and specifying the assumptions that we make.

In \code{S.formula}, we provide the variable names of $Z$ and $D$ on the left hand side, and the covariates of S-model on the right hand side. This specifies the multinomial model 
\begin{equation}
    \Pr(S = s \mid X_1, X_2) \propto \exp(\alpha_s + \beta_{1, s}X_1 + \beta_{2, s}X_2),
\end{equation}
where $X_1$ and $X_2$ refers to the age and COPD respectively, $\alpha_s$, $\beta_{1,s}$ and $\beta_{2,s}$ are parameters. Similarly, in \code{Y.formula}, we provide the variable name of $Y$ on the left hand side, and covariates of Y-model on the right hand side. This combined with \code{Y.family} specifies the outcome model as a generalized linear model \begin{align}
    Y \mid S = s, Z = z, X_1, X_2 \sim \mathrm{Ber}(\mu_{s,z}(X_1, X_2)), \\
    \mathrm{logit}\,\mu_{s,z}(X_1, X_2) = \gamma_{s,z} + \delta_{1,s,z}X_1 + \delta_{1,s,z}X_2,
\end{align} where $\gamma_{s,z}$, $\delta_{1,s,z}$ and $\delta_{2,s,z}$ are parameters.

We assume the existence of three possible strata, defined by $n = (0, 0)$, $c = (0, 1)$ and $a = (1, 1)$. We put an asterisk for $n = (0, 0)$ and $a = (1, 1)$ because we assume ER for these strata, which implies 
\begin{align}
    (\gamma_{n, 0}, \delta_{1, n, 0}, \delta_{2, n, 0}) = (\gamma_{n, 1}, \delta_{1, n, 1}, \delta_{2, n, 1}), \\
    (\gamma_{a, 0}, \delta_{1, a, 0}, \delta_{2, a, 0}) = (\gamma_{a, 1}, \delta_{1, a, 1}, \delta_{2, a, 1}).
\end{align}

Finally, we specify prior distributions for the intercepts $\alpha_s$, $\gamma_{s,z}$ and coefficients $\beta_{1,s}$, $\beta_{2,s}$, $\delta_{1,s,z}$ and $\delta_{2,s,z}$. In this case, we assume the prior distributions of all these parameters are standard normal distribution. 

\subsubsection{Fit the Principal Stratification Model}

We fit the model using the function \code{PStrata}.

\begin{CodeChunk}
\begin{CodeInput}
R> fit <- PStrata(
+    PSobj,
+    warmup = 1000, iter = 2000,
+    cores = 6, chains = 6
+  )
\end{CodeInput}
\end{CodeChunk}

In the above snippet, we fit the Bayesian principal stratification model by sampling 6 chains, each containing 2000 iterations including 1000 warm up iterations. The sampling job is distributed to 6 cores for efficiency. These parameters are directly passed to \code{rstan::stan()}, and detailed information can be found in the manual of \pkg{rstan}.

\begin{CodeChunk}
\begin{CodeInput}
R> summary(fit)
\end{CodeInput}
\begin{CodeOutput}
         mean         sd      2.5%       25%    median       75%     97.5%
  n 0.6940673 0.01196718 0.6707998 0.6859948 0.6940854 0.7022795 0.7170083
  c 0.1137423 0.01621615 0.0803734 0.1034171 0.1138241 0.1246476 0.1455330
  a 0.1921904 0.01055071 0.1717857 0.1850637 0.1918401 0.1992098 0.2128891
\end{CodeOutput}
\end{CodeChunk}

From the fitted model, 69.4\% (CI: 67.1\% to 71.7\%) of the patients are never-vaccinated, 19.2\% (CI: 17.2\% to 21.3\%) of the patients are always-vaccinated, and the remaining 11.4\% (CI: 8.0\% to 14.6\%) are compliant-vaccinated.

\subsubsection{Predict Mean Effects}

\begin{CodeChunk}
\begin{CodeInput}
R> outcome <- PSOutcome(fit)
R> summary(outcome, "matrix")
\end{CodeInput}
\begin{CodeOutput}
        mean       sd     2.5%      25%   median      75%    97.5%
n:0 0.081546 0.007444 0.067090 0.076531 0.081497 0.086451 0.096345
n:1 0.081546 0.007444 0.067090 0.076531 0.081497 0.086451 0.096345
c:0 0.166652 0.063261 0.060931 0.122136 0.160119 0.204622 0.310435
c:1 0.068589 0.027496 0.024559 0.048624 0.065305 0.085192 0.129120
a:0 0.100001 0.014312 0.073536 0.090090 0.099535 0.109130 0.129424
a:1 0.100001 0.014312 0.073536 0.090090 0.099535 0.109130 0.129424
\end{CodeOutput}
\end{CodeChunk}

We use \code{PSOutcome} to calculate the mean effect given stratum $s$ and treatment arm $z$, defined as
\begin{equation}
    \mu_{s,z} = \mathbb{E}[Y(z) \mid S = s].
\end{equation}
From the output, one can see that the mean outcome for both arms is 0.082 (CI: 0.067 to 0.096) for the never-vaccinated, and 0.100 (CI: 0.074 to 0.129) for the always-vaccinated. For these two strata, the mean outcomes for the treatment arm and the control arm are the same, which is a consequence of the ER assumption. For the compliant-vaccinated, the mean outcome is 0.167 (CI: 0.061 to 0.310) for the control arm and 0.069 (CI: 0.026 to 0.129) for the treatment arm.

\subsubsection{Calculate Contrasts}

To identify the difference in the mean outcome between the treatment and control arm, one can use the \code{PSContrast} function.

\begin{CodeChunk}
\begin{CodeInput}
R> contrast <- PSContrast(outcome, Z = TRUE)
R> summary(contrast, "matrix")
\end{CodeInput}
\begin{CodeOutput}
               mean       sd      2.5%       25%    median       75%    97.5%
n:{1}-{0}  0.000000 0.000000  0.000000  0.000000  0.000000  0.000000 0.000000
c:{1}-{0} -0.098063 0.068129 -0.246702 -0.140799 -0.092666 -0.052169 0.022559
a:{1}-{0}  0.000000 0.000000  0.000000  0.000000  0.000000  0.000000 0.000000
\end{CodeOutput}
\end{CodeChunk}

In the above snippet, \code{Z = TRUE} specifies that we would like to find contrasts between all levels of treatment $Z$. The contrasts for never-vaccinated and always-vaccinated are identically zero due to the ER assumption. The contrast for the compliers, also known as the complier average causal effect, is -0.098 (CI: -0.247, 0.023). 

\subsubsection{Conclusion}

From the principal stratification analysis, on average, encouragement on flu vaccination would reduce the average proportion of hospital visits from 16.7\% to 6.9\% for compliant-vaccinated. The difference is not statistically significant at 95\% level, but the confidence interval covers zero only by a tiny margin. The principal stratification analysis also managed to identify the difference in these subpopulations defined by the principal strata. For example, for the never-vaccinated, the average proportion of hospital visits is 8.2\%, while for the always-vaccinated, the average proportion of hospital visits is counterintuitively higher at 10.0\%. This might be due to the fact that those who do not get vaccination anyway consider themselves as low risk, so they have a lower chance of developing severe symptoms even without vaccination.

\section{Simulation} \label{sec:simulation}

In this section, we conduct simulation studies to illustrate how to use the package under different scenarios.

\subsection{Case 1: Two-sided noncompliance}
This simulation study features a classic non-compliance scenario where defiers ($S = (1, 0)$) are excluded from the analysis, with ER assumed within both always-takers ($S = (1,1)$) and never-takers ($S = (0, 0)$).

We generate 10000 units. Two covariates $X_1, X_2$ are independently sampled from $\mathcal{N}(0,1)$. We assign $S \in \{(0, 0), (0, 1), (1, 1)\}$ independently with probability 0.3, 0.5 and 0.2, respectively. The treatment assignment is random to every unit with $p(Z = 1) = 0.5$ to mimic a random clinical trial. Conditional on the stratum $S$ and treatment assignment $Z$, the outcome $Y$ follows a gaussian distribution where $$\begin{aligned}
  Y\mid S = (0,0), Z = z, X_1, X_2 &\sim \mathcal{N}(X_1 - X_2 + X_1X_2, 0.3)\\
  Y\mid S = (0,1), Z = z, X_1, X_2 &\sim \mathcal{N}(2X_1 - (1 + z)X_2 + 2 + 6z, 0.2)\\
  Y\mid S = (1,1), Z = z, X_1, X_2 &\sim \mathcal{N}(X_1 + X_2 - 1, 0.2).
\end{aligned}$$

We run the sampler with 6 chains and 1000 warmup iterations and 1000 sampling iterations for each chain. The true values of stratum probability and the causal effects and the respective posterior distributions are given in Table \ref{tbl::sim1}.

\begin{CodeChunk}
\begin{CodeInput}
R> fit <- PStrata(
+    S.formula = Z + D ~ 1,
+    Y.formula = Y ~ X1 * X2,
+    Y.family = gaussian(),
+    data = data,
+    strata = c(n = "00*", c = "01", a = "11*"),
+    prior_intercept = prior_normal(0, 1),
+    warmup = 1000, iter = 2000,
+    cores = 6, chains = 6, refresh = 10
+  )
R> summary(fit)
\end{CodeInput}
\begin{CodeOutput}
         mean          sd      2.5%       25%    median       75%     97.5%
  n 0.2888397 0.004491325 0.2800518 0.2858500 0.2888168 0.2918257 0.2976134
  c 0.5046871 0.004968135 0.4948485 0.5014518 0.5046069 0.5079710 0.5145919
  a 0.2064732 0.003959123 0.1988303 0.2037680 0.2064324 0.2091346 0.2143214
\end{CodeOutput}
\begin{CodeInput}
R> outcome <- PSOutcome(fit)
R> contrast <- PSContrast(outcome, Z = TRUE)
R> summary(contrast, "matrix")
\end{CodeInput}
\begin{CodeOutput}
              mean       sd     2.5%      25%  median      75%    97.5%
n:{1}-{0} 0.000000 0.000000 0.000000 0.000000 0.00000 0.000000 0.000000
c:{1}-{0} 5.992352 0.005649 5.981134 5.988609 5.99229 5.996191 6.003188
a:{1}-{0} 0.000000 0.000000 0.000000 0.000000 0.00000 0.000000 0.000000
\end{CodeOutput}
\end{CodeChunk}

\begin{table}[h]
  \centering
  \begin{tabular}{ccccc}
    \toprule
    & True value & Posterior mean & $2.5\%$ quantile & $97.5\%$ quantile \\
    \midrule
    $\mathrm{Pr}(S = (0,0))$ & 0.300 & 0.289 & 0.280 & 0.298 \\
    $\mathrm{Pr}(S = (0,1))$ & 0.500 & 0.505 & 0.495 & 0.515\\
    $\mathrm{Pr}(S = (1,1))$ & 0.200 & 0.206 & 0.199 & 0.214\\
    \midrule
    $\E[Y(1) - Y(0) \mid S = (0,1)]$ & 6.000 & 5.992 & 5.981 & 6.003\\
    \bottomrule
    \end{tabular}
    \caption{The posterior summary of important quantities with true values in Simulation 1.}
    \label{tbl::sim1}
\end{table}

\subsection{Case 2: Two post-treatment variables}

This simulation study features a more complex scenario where two post-treatment covariates $D_1$ and $D_2$ exist. Let the principal strata be defined by $S = (D_1(0), D_1(1), D_2(0), D_2(1))$. In this study, we include five out of 16 strata, namely $\mathcal{S} = \{0000, 0001, 0101, 0011, 1111\}$, and assume ER for 0000, 0011 and 1111. We do not include baseline covariates in this study.

We simulate 10,000 data for better identification of the strata. We randomly assign principal stratum $S$ and treatment status $Z$ to each unit, with the stratum-assignment probability being $p = (0.15, 0.3, 0.2, 0.2, 0.15)$ and the treatment assignment probability $P(Z = 1) = 0.5$. The outcome variable $Y$ is sampled from a Gaussian distribution given in Table \ref{tbl::outcome_sim2}.

\begin{table}[h]
  \centering
  \begin{tabular}{cccc}
    \toprule
    & Probability & $Z = 0$ & $Z = 1$ \\
    \midrule
    $S = 0000$ & 0.15 & \multicolumn{2}{c}{$\mathcal{N}(3,1)$} \\
    $S = 0001$ & 0.30& $\mathcal{N}(-1, 0.5)$ & $\mathcal{N}(-3, 0.5)$ \\
    $S = 0011$ & 0.20 & $\mathcal{N}(2, 0.5)$ & $\mathcal{N}(5, 0.5)$ \\
    $S = 0101$ & 0.20 & \multicolumn{2}{c}{$\mathcal{N}(-1, 3)$} \\
    $S = 1111$ & 0.15& \multicolumn{2}{c}{$\mathcal{N}(1, 2)$} \\
    \bottomrule
  \end{tabular}
  \caption{Outcome specification for Simulation 2}
  \label{tbl::outcome_sim2}
\end{table}

We run the sampler with 6 chains and 1000 warmup iterations and 1000 sampling iterations for each chain. The true values of parameters and the respective posterior means are given in Table \ref{tbl::sim2}. 

\begin{CodeChunk}
\begin{CodeInput}
R> fit <- PStrata(
+    S.formula = Z + D ~ 1,
+    Y.formula = Y ~ 1,
+    Y.family = gaussian(),
+    data = data,
+    strata = c(
+      nn = "00|00*", nc = "00|01", cc = "01|01", na = "00|11*", aa = "11|11*"
+    ),
+    prior_intercept = prior_normal(0, 1),
+    warmup = 1000, iter = 2000,
+    cores = 6, chains = 6
+  )
R> summary(fit)
\end{CodeInput}
\begin{CodeOutput}
          mean          sd      2.5%       25%    median       75%     97.5%
  nn 0.1536845 0.006951125 0.1412590 0.1481628 0.1537814 0.1591534 0.1663185
  nc 0.2784839 0.023542175 0.2465955 0.2555127 0.2798529 0.3014315 0.3093388
  cc 0.2263757 0.026944918 0.1922943 0.1997387 0.2288498 0.2529571 0.2599586
  na 0.1954942 0.005190111 0.1855834 0.1918809 0.1954155 0.1990191 0.2056500
  aa 0.1459617 0.003763111 0.1386817 0.1434026 0.1459706 0.1484749 0.1533925
\end{CodeOutput}
\begin{CodeInput}
R> outcome <- PSOutcome(fit)
R> summary(outcome, "matrix")
\end{CodeInput}
\begin{CodeOutput}
          mean       sd      2.5%       25%    median       75%     97.5%
nn:0  3.026044 0.037910  2.952770  3.000045  3.025884  3.052389  3.099262
nn:1  3.026044 0.037910  2.952770  3.000045  3.025884  3.052389  3.099262
nc:0  0.522369 1.518245 -1.016887 -0.995661  0.503381  2.039995  2.077103
nc:1 -2.981742 0.016428 -3.014958 -2.992721 -2.981645 -2.970739 -2.949451
cc:0  0.503428 1.501155 -1.018944 -0.997362  0.490139  2.004243  2.040805
cc:1  5.008498 0.018250  4.971816  4.996631  5.008637  5.021135  5.043137
na:0 -1.081677 0.075050 -1.227409 -1.133591 -1.081889 -1.031386 -0.933170
na:1 -1.081677 0.075050 -1.227409 -1.133591 -1.081889 -1.031386 -0.933170
aa:0  0.940070 0.062138  0.817387  0.897509  0.940663  0.982155  1.061605
aa:1  0.940070 0.062138  0.817387  0.897509  0.940663  0.982155  1.061605
\end{CodeOutput}
\begin{CodeInput}
R> contrast <- PSContrast(outcome, Z = TRUE)
R> summary(contrast, "matrix")
\end{CodeInput}
\begin{CodeOutput}
               mean       sd      2.5%       25%    median       75%     97.5%
nn:{1}-{0}  0.00000 0.000000  0.000000  0.000000  0.000000  0.000000  0.000000
nc:{1}-{0} -3.50411 1.520898 -5.072012 -5.024450 -3.497351 -1.983469 -1.949683
cc:{1}-{0}  4.50507 1.496066  2.963890  3.009801  4.513690  6.000999  6.035906
na:{1}-{0}  0.00000 0.000000  0.000000  0.000000  0.000000  0.000000  0.000000
aa:{1}-{0}  0.00000 0.000000  0.000000  0.000000  0.000000  0.000000  0.000000
\end{CodeOutput}
\end{CodeChunk}

\begin{table}[h]
  \centering
  \begin{tabular}{ccccc}
    \toprule
    & True value & Posterior mean & $2.5\%$ quantile & $97.5\%$ quantile \\
    \midrule
    $\mathrm{Pr}(S = 0000)$ & 0.150 & 0.153 & 0.141 & 0.166 \\
    $\mathrm{Pr}(S = 0001)$ & 0.300 & 0.278 & 0.247 & 0.309\\
    $\mathrm{Pr}(S = 0011)$ & 0.200 & 0.226 & 0.192 & 0.260\\
    $\mathrm{Pr}(S = 0101)$ & 0.200 & 0.195 & 0.186 & 0.206\\
    $\mathrm{Pr}(S = 1111)$ & 0.150 & 0.146 & 0.139 & 0.153\\
    \midrule
    $\E[Y(1) - Y(0) \mid S = 0001]$ & -2.000 & -3.504 & -5.072 & -1.950\\
    $\E[Y(1) - Y(0) \mid S = 0101]$ & 3.000 & 4.505 & 2.964 & 6.046\\
    \bottomrule
    \end{tabular}
    \caption{The posterior summary of important quantities with true values in Simulation 2.}
    \label{tbl::sim2}
\end{table}


\subsection{Case 3: Noncompliance with time-to-event outcome}
We simulate a randomized experiment with $N = 10,000$ units, under the non-compliance setting.  We assign $S \in \{(0, 0), (0, 1), (1, 1)\}$ independently with probability 0.25, 0.60 and 0.15, respectively. The treatment assignment $Z_i$ is independently drawn from a $Bernoulli(0.5)$. The true uncensored failure time is generated from the Weibull-Cox model \eqref{eq:Weibull-Cox}, with separate parameters for each of the six combination of stratum and the treatment assignment. We include not covariates and with an intercept, the Weibull-Cox model takes the form of 
\begin{equation}
    h(t; S_i = s, Z_i = z) = t^{\exp(\theta_{s,z}) - 1}\exp(\alpha_{s,z}).
\end{equation}
Table \ref{tab:sim-Tmodel} presents the T-model parameters for these two scenarios. We draw the censoring time $C_i$ independently from an exponential distribution with rate 0.3, leading to an event rate being approximately 35\%. 

\begin{table}[h]
    \centering
    \begin{tabular}{cccccc}
    \toprule
     \cmidrule(lr){3-6}
     & & \multicolumn{2}{c}{$\Pr(Y(0)\mid S=s)$} &  \multicolumn{2}{c}{$\Pr(Y(1)\mid S=s)$} \\ 
     \cmidrule(lr){3-4}\cmidrule(lr){5-6}
      $s$ & $\Pr(S = s)$ & $\theta_{s,0}$ & $\alpha_{s,0}$ & $\theta_{s,1}$ & $\alpha_{s,1}$ \\
      \midrule
      $n$ & 0.25 & $\log(2.0)$ & -4.0 & $\log(1.5)$ & -3.0 \\
      $c$ & 0.60 & $\log(1.5)$ & -2.5 & $\log(1.0)$ & -1.5 \\
      $a$ & 0.15 & $\log(1.0)$ & -1.0 & $\log(0.6)$ & 0.0 \\
      \bottomrule
    \end{tabular}
    \caption{Parameters of the true S-model and T-model in simulation 3.}
    \label{tab:sim-Tmodel}
\end{table}

\begin{CodeChunk}
\begin{CodeInput}
R> fit <- PStrata(
+    S.formula = Z + D ~ 1,
+    Y.formula = Y + delta ~ 1,
+    Y.family = survival(method = "Cox"),
+    data = data,
+    strata = c(n = "00", c = "01", a = "11"),
+    prior_intercept = prior_normal(0, 1),
+    warmup = 1000, iter = 2000,
+    cores = 6, chains = 6
+  )
R> summary(fit)
\end{CodeInput}
\begin{CodeOutput}
         mean          sd      2.5%       25%    median       75%     97.5%
  n 0.2337785 0.005976189 0.2221630 0.2296578 0.2337909 0.2378282 0.2453549
  c 0.6112725 0.007849057 0.5959362 0.6058670 0.6113284 0.6166199 0.6265693
  a 0.1549490 0.005075883 0.1451325 0.1514720 0.1549051 0.1585060 0.1649443
\end{CodeOutput}
\begin{CodeInput}
R> outcome <- PSOutcome(fit)
R> plot(outcome) + xlab("time") + ylab("survival probability") 
\end{CodeInput}
\end{CodeChunk}

\begin{figure}[htbp]
    \centering
    \includegraphics[width = 0.9\textwidth]{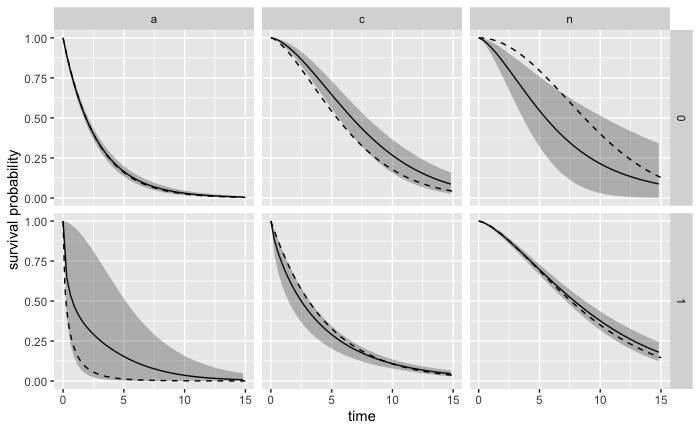}
    \caption{Plot of estimated survival probability curves (solid) with 95\% confidence region (shaded). The true survival probability curves are also plotted (dashed).}
    \label{fig:sim3plot}
\end{figure}

The estimated probabilities for three strata are respectively 0.234 (CI: 0.222 to 0.245), 0.611 (CI: 0.596 to 0.627) and 0.155 (CI: 0.145 to 0.165), close to the true values 0.25, 0.60 and 0.15. With \code{plot()}, we can view the estimated survival probability curves and their $95\%$ confidence regions visually (Figure \ref{fig:sim3plot}. The true survival probability curves are added for reference.

\subsection{Case 4: Data with cluster effects}

We simulate a randomized experiment with $N = 1,000$ units. We assign $S \in \{(0, 0), (0, 1), (1, 1)\}$ independently with probability 0.25, 0.50 and 0.25, respectively. The treatment assignment $Z_i$ is independently drawn from a $Bernoulli(0.5)$. To mimic the existence of clusters, we randomly assign these units to 10 clusters, denoted by $C_i = j$. We sample random effect $\xi_j$ from the standard normal distribution for each cluster $j$. Conditional on the stratum $S$ and treatment assignment $Z$, the outcome $Y$ follows a gaussian distribution where $$\begin{aligned}
  Y\mid S = (0,0), Z = z, X_1, X_2, C = j &\sim \mathcal{N}(X_1 - X_2 + \xi_j, 0.3)\\
  Y\mid S = (0,1), Z = z, X_1, X_2, C = j &\sim \mathcal{N}(2X_1 - (1 + z)X_2 + 2 + 6z + \xi_j, 0.2)\\
  Y\mid S = (1,1), Z = z, X_1, X_2, C = j &\sim \mathcal{N}(X_1 + X_2 - 1 + \xi_j, 0.2).
\end{aligned}$$
We run the sampler with 6 chains and 1000 warmup iterations and 1000 sampling iterations for each chain. The true values of stratum probability and the causal effects and the respective posterior distributions are given in Table \ref{tbl::sim4}.

\begin{CodeChunk}
\begin{CodeInput}
R> fit <- PStrata(
+    S.formula = Z + D ~ 1,
+    Y.formula = Y ~ X1 + X2 + (1 | C),
+    Y.family = gaussian(),
+    data = data,
+    strata = c(n = "00*", c = "01", a = "11*"),
+    prior_intercept = prior_normal(0, 1),
+    warmup = 1000, iter = 2000,
+    cores = 6, chains = 6, refresh = 10
+  )
R> summary(fit)
\end{CodeInput}
\begin{CodeOutput}
         mean         sd      2.5%       25%    median       75%     97.5%
  n 0.2806836 0.01412599 0.2520836 0.2711908 0.2800077 0.2902168 0.3083786
  c 0.5202746 0.01656015 0.4871375 0.5084747 0.5208637 0.5312158 0.5520182
  a 0.1990418 0.01242839 0.1757518 0.1901582 0.1988140 0.2076523 0.2224745
\end{CodeOutput}
\begin{CodeInput}
R> outcome <- PSOutcome(fit)
R> contrast <- PSContrast(outcome, Z = TRUE)
R> summary(contrast, "matrix")
\end{CodeInput}
\begin{CodeOutput}
              mean        sd     2.5%      25%  median      75%    97.5%
n:{1}-{0} 0.000000 0.0000000 0.000000 0.000000 0.00000 0.000000 0.000000
c:{1}-{0} 5.977509 0.0178666 5.943497 5.965294 5.97707 5.989835 6.012452
a:{1}-{0} 0.000000 0.0000000 0.000000 0.000000 0.00000 0.000000 0.000000
\end{CodeOutput}
\end{CodeChunk}

\begin{table}[h]
  \centering
  \begin{tabular}{ccccc}
    \toprule
    & True value & Posterior mean & $2.5\%$ quantile & $97.5\%$ quantile \\
    \midrule
    $\mathrm{Pr}(S = (0,0))$ & 0.300 & 0.281 & 0.252 & 0.308 \\
    $\mathrm{Pr}(S = (0,1))$ & 0.500 & 0.520 & 0.487 & 0.552\\
    $\mathrm{Pr}(S = (1,1))$ & 0.200 & 0.199 & 0.176 & 0.222\\
    \midrule
    $\E[Y(1) - Y(0) \mid S = (0,1)]$ & 6.000 & 5.978 & 5.943 & 6.012\\
    \bottomrule
    \end{tabular}
    \caption{The posterior summary of important quantities with true values in Simulation 4.}
    \label{tbl::sim4}
\end{table}

\section{Summary} \label{sec:summary}

\PS is an important tool for causal inference with post-treatment confounding. This paper introduces the \pkg{PStrata} package and demonstrates its use in the most common settings of non-compliance with single and two post-treatment variables. \pkg{PStrata} is under continuing development; future versions will include the settings of truncation by death and informative missing data, as well as graphic diagnostics. 

\section*{Computational details}
\pkg{PStrata} 0.0.1 was built on \proglang{R}~4.0.3 and dependent on 
the \pkg{Rcpp}~1.0.6 package, the \pkg{dplyr}~1.0.7 package, the \pkg{purrr}~0.3.4 package, the \pkg{rstan}~2.21.1 package, the \pkg{lme4}~1.1-27.1 package, the \pkg{ggplot2}~3.3.6 package and the \pkg{patchwork}~1.1.1 package. All these packages used are available from the Comprehensive \proglang{R} Archive Network (CRAN) at
\url{https://CRAN.R-project.org/}.

\section*{Acknowledgement}
This research is supported by the Patient-Centered Outcomes Research Institute (PCORI) contract ME-2019C1-16146. We thank Laine Thomas and Peng Ding for helpful comments.


\bibliography{PStrata}

\end{document}